\documentclass[sigconf]{acmart} 

\usepackage{caption}
\usepackage{subcaption}
\usepackage{multirow}
\usepackage{booktabs}
\usepackage{todonotes}

\AtBeginDocument{%
  }

\acmConference[CHASE '23]{IEEE/ACM International Conference on Connected Health: Applications, Systems and Engineering Technologies}{June 21--23, 2023}{Orlando, FL}

\acmSubmissionID{40}
\begin{document}

\title{Social Visual Behavior Analytics for Autism Therapy of Children Based on Automated Mutual Gaze Detection}


\author{Zhang Guo}
\affiliation{%
  \institution{University of Delaware}
  \city{Newark}
  \state{DE}
  \country{United States}
}
\email{guozhang@udel.edu}
\orcid{0000-0003-1599-571X}

\author{Vuthea Chheang}
\affiliation{%
  \institution{University of Delaware}
  \city{Newark}
  \state{DE}
  \country{United States}
}
\email{Vuthea@udel.edu}
\orcid{0000-0001-5999-4968}

\author{Jicheng Li}
\affiliation{%
  \institution{University of Delaware}
  \city{Newark}
  \state{Delaware}
  \country{United States}
}
\email{lijichen@udel.edu}
\orcid{0000-0003-2564-6337}

\author{Kenneth E.Barner}
\affiliation{%
  \institution{University of Delaware}
  \city{Newark}
  \state{Delaware}
  \country{United States}
}
\email{barner@udel.edu}
\orcid{}

\author{Anjana Bhat}
\affiliation{%
  \institution{University of Delaware}
  \city{Newark}
  \state{DE}
  \country{United States}
}
\email{abhat@udel.edu}
\orcid{0000-0002-0771-0967}

\author{Roghayeh Barmaki}
\authornote{Correspondence: Roghayeh Barmaki (rlb@udel.edu)}
\affiliation{%
  \institution{University of Delaware}
  \city{Newark}
  \state{DE}
  \country{United States}
}
\email{rlb@udel.edu}
\orcid{0000-0002-7570-5270}

\renewcommand{\shortauthors}{Guo et al.}
\newcommand{\etal}{\textit{et al. }}

\begin{abstract}
  
  Social visual behavior, as a type of non-verbal communication, plays a central role in studying social cognitive processes in interactive and complex settings of autism therapy interventions.
  However, for social visual behavior analytics in children with autism, it is challenging to collect gaze data manually and evaluate them because it costs a lot of time and effort for human coders.
  In this paper, we introduce a social visual behavior analytics approach by quantifying the mutual gaze performance of children receiving play-based autism interventions using an automated mutual gaze detection framework.
  Our analysis is based on a video dataset that captures and records social interactions between children with autism and their therapy trainers ($N\!=\!28$ observations, $84$ video clips, $21$ Hrs duration).
  The effectiveness of our framework was evaluated by comparing the mutual gaze ratio derived from the mutual gaze detection framework with the human-coded ratio values.
  We analyzed the mutual gaze frequency and duration across different therapy settings, activities, and sessions.
  We created mutual gaze-related measures for social visual behavior score prediction using multiple machine learning-based regression models.     
  The results show that our method provides mutual gaze measures that reliably represent (or even replace) the human coders' hand-coded social gaze measures and effectively evaluates and predicts ASD children's social visual performance during the intervention.
  Our findings have implications for social interaction analysis in small-group behavior assessments in numerous co-located settings in (special) education and in the workplace.
  
\end{abstract}



\begin{teaserfigure}
    \centering
    \includegraphics[width=\textwidth]{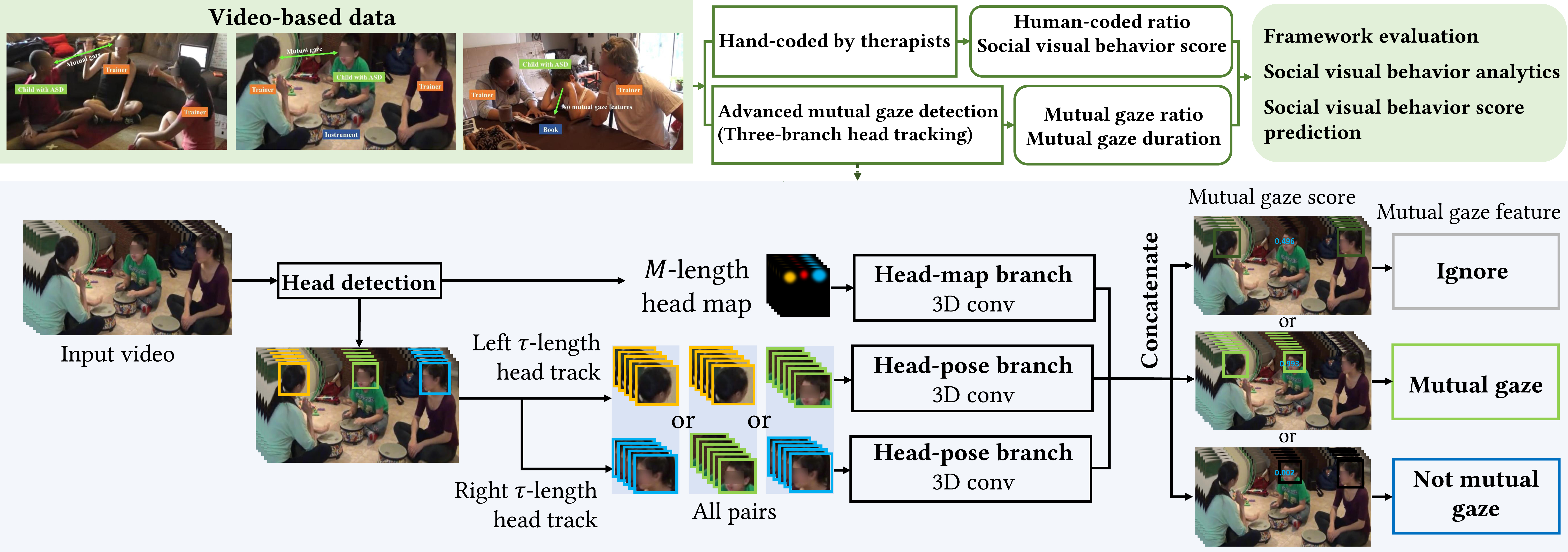}
    \caption{Overview of the mutual gaze detection and social visual behavior analytics for children with autism based on a video-based ASD dataset collected from a physical therapy intervention (top) and the deep learning architecture for mutual gaze detection, adopted from the three-branch head tracking framework~\cite{9311655} (bottom).
    }        
    \label{fig:1}
\end{teaserfigure}

\maketitle
\vspace{20pt}
\section{Introduction}
\label{Sec:Intro}

As a type of non-verbal communication, social visual behavior plays a central role in studying social cognitive processes in interactive and complex settings.
Visual cues are mostly displayed and perceived spontaneously by humans~\cite{akhtar2017visual}, and it is commonly believed that the level of cognitive abilities and social skills are reflected in gaze behavior~\cite{capozzi2018attention}.
More specifically, 
individuals and children with Autism Spectrum Disorder (ASD) 
have difficulty in identifying, performing, and maintaining such gaze behaviors in social interactions and communication~\cite{zablotsky2019prevalence}.
The lack of such social engagement may lead to anxiety, depression, and social avoidance in individuals on the autism spectrum during social interactions~\cite{mavadati2014comparing}. 

Individuals on the autism spectrum have a poor quality of social interactions, e.g., reduced eye contact~\cite{dawson2004early}, reduced interest in social stimuli~\cite{ozonoff2010prospective}, lack of response to name~\cite{osterling2002early}, and insufficient sharing of interests~\cite{yoder2009predicting}), and lack of social connections with partners~\cite{sterling2008characteristics}.
Mutual gaze, or two people looking at each other, is an essential type of social gaze ~\cite{dawson2004early} and has been considered a critical indicator for establishing and maintaining successful face-to-face interactions during everyday interactions~\cite{palmero2018automatic}.
In many autism therapy interventions, mutual gaze has been used as an objective measure to interpret the social behaviors of children with autism and evaluate therapy effectiveness ~\cite{thepsoonthorn2015look,richardson2005looking,guo2021automated}.

In this paper, we introduce a social visual behavior analytics approach for observing the mutual gaze performance of children with autism during therapy sessions using mutual gaze detection (see Figure~\ref{fig:1}). 
This work uses data collected from a play therapy intervention for children with autism ($N\!=\!21$ children, $7.57\pm 2.31$ of years old).
The training sessions were captured by a standard camera and recorded as video clips.
We enhanced the detection technique \cite{guo2021automated} and automatically extracted the mutual gaze features from the $84$ video recording sessions of children using an advanced mutual gaze detection~\cite{9311655}, and we generated the social visual behavior measures based on the framework outcomes.

The effectiveness of our framework was assessed by comparing the mutual gaze ratio generated by the framework outcomes with hand-coded ratio measures annotated by human experts.
The social-visual behavior of children with autism was examined across different therapy settings, training activities, and therapy sessions.
We also predicted the social visual behavior score using multiple machine learning-based regression models with our mutual gaze-related measures.
We found that the random forest model achieved the best performance, assisted by the level of functioning skills and social affect score from the Autism Diagnostic Observation Schedule (ADOS)~\cite{lord2012autism}.

Our main contributions are the following:
\begin{itemize}
    \item Providing an innovative application in autism therapy of mutual gaze utilizing a detection framework and deep learning.
    \item Generating objective indicators to quantify social visual behaviors with the ASD dataset for fine-grained measures and balanced analysis.
    \item Introducing a new social gaze analytics approach using various learning analytics approaches and multiple machine learning-based regression models to analyze, evaluate, and predict social behaviors in various ASD therapy settings.
\end{itemize}
 

\section{Related Work} 
\label{Sec:RelatedWork}

In the following, we describe related work and background on visual behavior in children with autism and mutual gaze detection.

\subsection{Visual Behavior in Children with Autism}



Children with autism are immediately noticeable for their reduced eye contact and eye-to-face gaze when expected to engage in reciprocal behaviors~\cite{mirenda1983gaze}.
Visual behavior analytics has drawn considerable attention for its ability to offer dynamic insights into emotions and social abilities in individuals with autism~\cite{rutter1978diagnosis,wing1979severe,minissi2021assessment}.
Although research into gaze behavior in autism individuals has identified some general patterns, it has also yielded some inconsistent findings~\cite{hessels2018eye,guillon2014visual,buitelaar1995attachment}: some studies ~\cite{falck2011special,chita2016social} using pictures and videos suggest that individuals with autism avoid looking directly into others' eyes, whereas others indicate that they have typical gaze patterns as control groups~\cite{frazier2017meta}.
It has been illustrated that some of these discrepancies are probably due to the wide spectrum in autism~\cite{canigueral2019role}, the lack of experimental paradigms for studying eye gaze in real social interaction~\cite{chevallier2015measuring,von2017high}, and lack of consistency among the conducted experimental settings and analysis approaches~\cite{volkmar1990gaze,drysdale2018gaze,canigueral2019role}.


There has been a growing interest in visual behavior analytics in autism during face-to-face interaction. 
Volkmar \etal~\cite{volkmar1990gaze} collected gaze patterns of $20$ individuals with autism in educational settings as they interacted with other staff. 
They reported that people with autism attended less to staff during one-on-one interactions than the typical controls.
In Zhao's work ~\cite{zhao2021characteristics}, they compared the gaze behavior of $20$ children with autism to $23$ typically developing children during face-to-face conversations. 
Experiments show that children with autism significantly preferred to look at the background than facial areas, and 
they concluded that gaze patterns varied depending on the conversational subject and how interesting it was to the children.
Furthermore, although evidence regarding interpersonal dynamics is mixed, it is generally accepted that individuals with autism have difficulty with gaze dynamics during social interactions ~\cite{dawson2004early,bonnel2003enhanced,chita2016social}.

\subsection{Mutual Gaze Detection}

Researchers have shown that tracing eye-to-face gaze from visual-based data offers a powerful mechanism for visual behavior analytics and assessment of the affective and cognitive behaviors of individuals with autism~\cite{zhao2021characteristics,chita2016social}.
In order to investigate how individuals with autism visually attend, early studies used hand-coding annotations by raters for gaze behavior~\cite{mundy1994joint}. 
Typically, manual coding was conducted frame by frame to determine the gaze allocation, which suffered from the
limitations of inaccuracy because of 
the rater’s subjective factors, and are time- and labor- intensive~\cite{boraston2007application}. 

With the advancements in hardware technology, wearable devices for gaze tracking can address these limitations by providing 
a sensitive and accurate measure of gaze allocation~\cite{ye2012detecting,ye2015detecting,alviar2020multimodal,rahal2019understanding}.
Although those devices may provide accurate measures 
for manual annotators and have been successfully applied in some real-world scenarios~\cite{hessels2020task,hessels2020looking}, they are expensive for mainstream applications and unsuitable for naturalistic or face-to-face interactions in large groups~\cite{funke2016eye,hosp2020remoteeye}. Moreover, the ultra sensitivity of children and individuals with autism prevents them from comfortably wearing these tracking gadgets due to sensory overload.

A practical alternative would be to estimate gaze direction and extract gaze features using deep learning methods~\cite{marin2014detecting,liaqat2021predicting}. 
Compared with eye trackers for single-feature extraction, cameras and computers are easier to access and have been widely used in daily life for multiple purposes, and deep learning methods can be applied for various feature extraction~\cite{minaee2021deep,gunes2007bi,heredia2021automatic}.
Researchers have used deep learning-based object detection and gaze-following approaches to identify shared attention features among students in previous works~\cite{guo2019collaboration,barmaki2020deep}. 

The issue of detecting mutual gaze behaviors in videos was also implemented in recent studies~\cite{marin2014detecting, palmero2018automatic, guo2021automated}.
Marin-Jimenez \emph{et al.} 
\cite{marin2014detecting} introduced a mutual gaze detection framework that computed the mutual gaze score between pairs of heads per frame by modeling and predicting pitch and yaw angles of the human heads with a 2D gaussian process regression model based on the detected heads.
Despite the effectiveness, this work only focused on single-frame detection and lost the consistency of adjacent frames.
Later, they provided a new method for mutual gaze detection, which could derive gaze information from a temporal period of head tracks~\cite{marin2019laeo}.

To increase the detection accuracy and enhance the automated detection technique~\cite{guo2021automated}, we adopt the advanced mutual gaze detection framework~\cite{9311655} in this paper, which automatically extracts mutual gaze features based on the head tracks with extended temporal dimension head maps and considers multiple consecutive frames, instead of single frames, in order to reduce the influence of noise, inconsistency, and detection problems.
The head maps also encode the relative 3D arrangement (depth) of the people in the scene.  

Compared with other detectors for mutual gaze, the deep learning-based framework we utilize~\cite{9311655} requires only a standard camera in a fixed position with one or more pairs of heads in the scene. 
It predicts mutual gaze scores for all possible pairs in the scene. It also handles the situation in which the head is not facing to the camera, or a third person cuts the gaze ray between the two-side people.
In this paper, our work not only provides an automated social visual behavior assessment tool but also provides a generalized social visual behavior analysis method using balanced data with more representative samples with less bias.



\section{Materials and Methods}
\label{Sec:Method}

In this section, we describe the details of our deep learning-based framework for mutual gaze extraction, provide the ASD dataset we collected, and introduce our measures, which we use for social visual behavior analysis.

\subsection{Data Collection in an Autism Therapy Intervention}
\label{Sec:Intervention}

In order to develop a social visual behavior analysis approach, particularly for autism therapy efficacy, we collected a number of video clips of various training conditions in an autism therapy intervention through a multi-session human subjects study.
The therapy intervention enrolled a total of $36$ children with autism in a randomized controlled trial study. 
This study was approved by the university's Institutional Review Board (IRB).
Children with autism were recruited through online and onsite flyers in local schools, self/parent advocacy groups, and services. Additionally, they were randomly assigned to one of the therapy groups.

The intervention lasted ten weeks, with pre-test and post-test sessions conducted in the first and last weeks, and autism therapy training activities provided twice a week during the intermediate eight weeks.
A pre-test was administered in the first week of treatment, and therapists assessed the children based on their pre-test performance in terms of social, communication, and functional abilities.
During the following eight weeks, they had two therapy sessions per week.
In the course of therapy sessions, two trainers assisted the child to engage in embodied creative activities in the home environment.
One trainer introduced the details of the activity and provided guidance, and the other trainer practiced the activity with the child as a model.
All the trainers, as pediatric physical therapists or graduate students/faculties, were well trained by autism therapists.
With the parents' permission and the notification to the children, the therapy sessions were videotaped with a standard camera, which was located at a fixed position toward the child.
Finally, the post-test session was completed by all children in the last week to examine changes in social, communication, and functional abilities.

We mainly focus, in this paper, on children's visual interaction with other people during the training activity. Therefore, two therapy groups with $28$ children were selected as a specific subset of video data (examples are shown in Figure~\ref{fig:1}).

\begin{itemize}

    \item \textbf{Play Therapy}:
    In the Play Therapy group, children were engaged in an improvisational "Music Making'' activity using musical instruments.
    A typical therapy session involves the child following the movements of the trainers, playing instruments, composing with the trainers, or singing a song along with hand clapping and waving gestures.
    \newline
    
    \item \textbf{Standard Therapy}:
    In the Standard Therapy group, children were engaged in a tabletop reading activity using age-appropriate books.
    During the therapy session, the trainer guided the child to follow instructions, read books, answer questions and express themselves spontaneously.
\end{itemize}

\noindent Play Therapy groups had one trainer seated opposite the child and one trainer seated next to the child, whereas both trainers of the Standard Therapy group sat next to the child.
Both groups engaged in social communication activities (i.e., eye contact, turn-taking, and non-verbal/verbal communication); however, the Standard Therapy group primarily focused on tabletop reading, whereas the Play Therapy group emphasized gross motor skills.

\paragraph{\textbf{ASD Dataset}}

We prepared an ASD dataset based on the video data collected from the autism therapy study described above. 
Given that data were collected in the wild without the therapists having any prior knowledge of the analysis methods and mutual gaze detection framework, a data balance issue occurred.
For our dataset, $7$ out of $28$ children's video records were excluded due to camera occlusion, low resolution, or incomplete data records or video annotations.
Each record included video clips from sessions $1$, $8$, and $16$, which covered the beginning, middle, and end of the therapy.
Other therapy sessions ($13$ video clips for each child record) were excluded from the analysis due to the lack of annotation. 

After post-processing, our ASD dataset had a total of $21$ hours of video from children in $84$ video clips ($30$ for the Standard Therapy group with ''Reading'' activity, $33$ for the Play Therapy group with ''Music Making'' activity, and $21$ for the Play Therapy group with ''Hello Song'' activity) with $21$ children (Standard Therapy: $10$, Play Therapy: $11$).
Seven children from the Play Therapy group did both the ''Music Making'' and the ''Hello Song'' activities.
Each video clip was approximately $15$ minutes with $25$ fps.
In this work, the ASD dataset was used as the input of the mutual gaze detection framework to calculate the children's mutual gaze score during the therapy session for further visual behavior analyses.

\paragraph{\textbf{Participants Profile Surveys}}
In order to confirm the eligibility of each child, parental consent was obtained, and the Social Communication Questionnaire (SCQ \cite{rutter2003scq}) and ADOS-2 measure ~\cite{lord2012autism} were completed before enrollment.
The survey data included the children's demographic information, including age, gender, ADOS-2 social affect score (collected before enrollment), and level of functioning measures (collected in pre-test sessions). 

In our ASD dataset, there were $21$ ($3$ females) children ranging in age from 5 to 12 years old.
Therapists rated each child's level of functioning on a scale of 1 to 3 on the basis of the child's level of independence in daily living skills ($1$: low functioning or needing significant support, $2$: medium functioning or needing moderate support, and $3$: high functioning or needing less support).
See Table~\ref{tab:t1} for the demographic characteristics of children in our ASD dataset.
Except for gender, the demographic information of children in our dataset is balanced for two therapy groups.

\begin{table*}[t]
    \centering
    \footnotesize
    \caption{Demographic characteristics of children in our ASD dataset.}
    \label{tab:t1}
        \begin{tabular}{lcccc}
        \toprule
        Participant Characteristics & Play Therapy & Standard Therapy & $F$ or $\chi^{2}$ value & $p$-value \\\midrule
        Age ($M \pm SD$) & $7.82 \pm 2.52$  & $7.30 \pm 2.16$ & $0.88$ & $0.55$\\
        Gender & $8$ M, $3$ F  & $10$ M, $0$ F & $3.18$ & $0.07$\\
        ADOS-2 social affect score ($M \pm SD$) & $17.27 \pm 4.56$  & $16.30 \pm 5.64$  & $1.88$ & $0.20$\\
        Level of Functioning ($M \pm SD$) & $1.91 \pm 0.83$ & $2.30 \pm 0.82$ & $0.58$ & $0.57$ \\
        \bottomrule
        \end{tabular}
        \begin{center}
        {\raggedright
        \centering
        \textit{
        The level of functioning skill is scored in a range from $1$ (low) to $3$ (high). \\
        Except for gender, the demographic information of children in our ASD dataset is balanced for two therapy groups.}
        \par}
        \end{center}
\end{table*}

\subsection{Mutual Gaze Detection Framework}

Using video recordings of therapy sessions, we aim to develop a deep learning-based framework that can automatically determine if there is any mutual gaze type of social visual interaction between the therapy trainers and the children.
The framework can potentially serve the therapists and trainers to observe the social state of the children during the entire training process.
To this end, we describe the development of our mutual gaze detection framework based on a state-of-the-art three-branch track network~\cite{9311655}.
Using the framework, we can retrieve mutual gaze features between all possible pairs of participants in the video (see Figure~\ref{fig:1} for more information about the framework).


\begin{figure*}[t]
    \centering
    \includegraphics[width=0.91\textwidth]{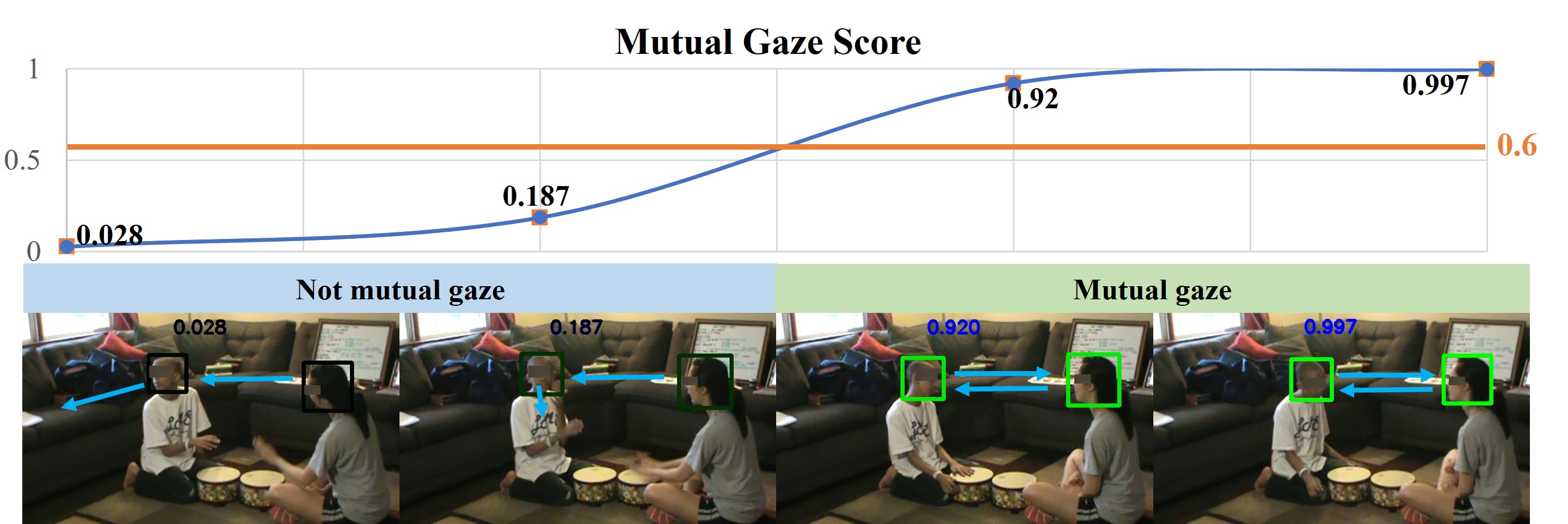}
    \caption{The results of mutual gaze detection for child-trainer pair in four sample frames. The mutual gaze score increases when the child is looking toward the trainer. 
    The cut-off point for mutual gaze features was set to be 0.6 since frames with values below this threshold do not display mutual gaze.
    Mutual gaze score is in the [$0$--$1$] range.}
    \label{fig:fwr}
    \Description{The mutual gaze detection results for child-and-trainer pair in four sample frames from one music group. The child is looking toward the trainer, and the score is increasing towards $1$. When the score is higher than $0.6$, the mutual gaze feature appears from the frame. A mutual gaze score curve is above the frames, and mutual gaze features are under the frames.}
\end{figure*}

Our framework focuses on detecting an individual's head pose (\emph{i.e.}, position and orientation) instead of relying on faces for gaze estimation, as there is no guarantee that the eyes will be fully visible in the collected scenes.
Firstly, the input video clip is sent to the Single Shot Multibox Detector (SSD~\cite{liu2016ssd}) for head detection.
Then the linking algorithm~\cite{singh2017online} groups them into tracks as the input of the three-branch track network.
The network has three branches, including two head-pose branches and one head-map branch (see Figure~\ref{fig:1}).
Each head-pose branch encodes a tensor of $\tau$ RGB frame crops of size 64$\times$64 pixels, which contains the head sequence of one person of the target pair, taking into account the head pose.
The head-map branch embeds relative head positions and relative distance to the camera (i.e., depth) between two head tracks over time using a 64$\times$64$\times$M map with 2D Gaussians for the whole $\tau$-frame track.

This framework extends the temporal dimension of the head maps and considers multiple consecutive frames instead of single frames in order to reduce the influence of noise, inconsistency, and detection problems.
The different Gaussian sizes of the head map are proportional to the head sizes (\emph{i.e.}, detected bounding boxes), which encodes the relative 3D arrangement (depth) of the people in the scene.  
In addition to the two head tracks, the position information for other persons in the scene is also encoded by the head-map branch in order to detect instances in which the third person cuts the gaze ray between the two-side people.

Each head-pose branch consists of five 3D convolutional layers, and the head-map branch consists of four 3D convolutional layers instead of the 2D ones in other works.
A fusion block is adopted to concatenate outcome embedding vectors from different branches after applying L2-normalization. 
It includes a fully connected layer, a dropout layer, and a Softmax layer that calculates a confidence score on the presence of eye contact between two people (mutual gaze).
It is important to note that the framework applies to all pairs of head tracks appearing simultaneously in video clips.

The framework yields mutual gaze scores for all valid pairs in frames.
In each frame, the shade of the detected bounding boxes of each pair is based on the confidence score (see Figure~\ref{fig:fwr}, for example).
The score shows how likely the pair of the target pair is to have eye contact interaction.
A higher score (lighter bounding boxes) indicates a greater probability of people looking at each other.
The framework ~\cite{9311655} achieves the state-of-the-art results on the TVHID ~\cite{patron2010high}, a benchmark dataset consisting of $300$ video clips from $20$ television shows representing five different types of human interaction. 
It is capable of detecting mutual gaze features under a variety of illuminations, scales, and cluttered background conditions.

We implemented the framework using TensorFlow~\cite{Martin2016Tensorflow}.
The framework was pre-trained using one Nvidia Tesla V100 GPU on two datasets:  AVA/UCO-LAEO dataset~\cite{9311655} including videos from $298$ movies and $4$ TV shows with annotated heads with bounding boxes, and AFLW dataset~\cite{koestinger2011annotated} containing about $25$k annotated faces in images.
During the test process, we set the head track length $\tau\!=\!10$ and head map length $M\!=\!10$. 
This framework pre-trained with AVA/UCO-LAEO dataset could achieve Average Precision ($AP$) $=\!92.3$\% ($M\!=\!1$) for the TVHID dataset and outperform the previous framework by 4--7\% for UCO-LAEO and around 18\% for AVA-LAEO~\cite{9311655}.

Since this paper focuses on the social visual behaviors of children as the main participants, the mutual gaze detection results between the trainers themselves were excluded from further analysis.
The framework automatically detected and generated mutual gaze scores in $10$--$20$ minutes for each video clip, which saves more time and effort than human annotation ($1$--$2$ hours for each clip).

\subsection{Measures}
\label{Sec:Measures}

In the therapy sessions, children looked in the direction of trainers and maintained their gaze towards the trainers while learning, playing, turn-taking, collaborating, and speaking to trainers during the activities.
To analyze the social visual behavior of children with autism in the therapeutic interventions, gaze interactions between the child and the trainers can be detected by our mutual gaze detection framework.
Here we describe the key measures used in our work.

\subsubsection{Measures Automatically Derived from Framework Outcomes}

\paragraph{\textbf{Mutual Gaze Ratio}}

Frame-based mutual gaze scores are generated from the outcomes of the mutual gaze detection framework.
The mutual gaze score of child-trainer pairs is defined as the possibility that they are looking at each other in the frame based on the detection of the track.
The frequency of the mutual gaze between child-trainer pairs over the course of the therapy sessions is defined as \emph{mutual gaze ratio}.
To calculate the mutual gaze ratio for each child, our framework counts the number of moments (frames) when the child and trainers had eye contact (frames that the mutual gaze score is higher than our threshold $0.6$ in the session $1$, $8$, and $16$), 
and divide it by the number of total session time (frames). See Figure~\ref{fig:fwr} for examples. 
Mutual gaze ratio is therefore normalized from $0$ (no mutual gaze at all) to $1$ (mutual gaze all the time).

\paragraph{\textbf{Mutual Gaze Duration}}

According to the mutual gaze scores provided by the mutual gaze detection framework outcome, we define \emph{mutual gaze duration} for each child based on the average duration of the mutual gaze between the child-trainer pairs throughout therapy sessions.
Our framework counts the moments (frames) when the child and trainers maintained eye contact for more than one second (in the session $1$, $8$, and $16$), \emph{e.g.}, the consecutive frames (more than $25$ frames) that the mutual gaze score is higher than the threshold $0.6$, and compute the mean value of the consecutive frame numbers for each child.

\subsubsection{Measures Hand-Coded by Human Experts}

\paragraph{\textbf{Human-Coded Ratio}}

While the measure of mutual gaze ratio is calculated by the framework automatically, we also have a measure of \emph{human-coded ratio} for the frequency of children's mutual gaze performance during the session $1$, $8$, and $16$, which is hand-coded by the human experts as the ground truth. 
Since the trainers performed interactively all the time and their gaze directions were toward the child for most of the time during the therapy sessions, the hand-coded mutual gaze annotation only focused on the children's social gaze behaviors.
The human-coded ratio is calculated by the time (second) the child looks at the trainers divided by the total time (second) of entire therapy sessions.
As a normalized measure of gaze interaction over entire sessions, the range of this measure is from $0$ (never looking at the trainers) to $1$ (looking at the trainers all the time).
This measure, as the ground truth, is compared with the mutual gaze ratio generated from the mutual gaze detection framework outcomes for framework evaluation.

\paragraph{\textbf{Social Visual Behavior Score}}

\emph{Social visual behavior score}, as a measure of children's looking patterns (looking at the trainers/objects/self or looking away), is hand-coded by the therapy experts.
There was one single social visual behavior score for each child, and therapy experts looked at the non-verbal interactive behavior attentiveness of the child during the clinician's single session interaction ($45$--$60$ mins) in the pre-test to score this measure on the scale of $1$ (looking away and lack of gaze interaction with trainers) to $4$ (frequent and sustained social gaze towards trainers to receive feedback) with $0.5$ intervals.

$100\%$ inter-rater reliability agreement was established between two coders for the children's looking patterns based on $20\%$ of video data.
Both human-coded mutual gaze ratio and social visual behavior score are derived from the annotated looking patterns.

\section{Results}
\label{Sec:Results}

In the following, we report descriptive and inferential statistics results about our measures described in the previous section.
We evaluate our framework by comparing the mutual gaze ratio automatically generated from the mutual gaze detection framework outcomes with the human-coded ratio annotated by human experts.
Then, we analyze children's social gaze behavior across different therapy groups, training activities, and therapy sessions.
Finally, we predict the social visual behavior score using different machine learning-based regression models.
In our results, significance was set at $p\!\leq\!0.05$.

\subsection{Framework Evaluation}

We first report descriptive and inferential statistics results about our measures described in the previous section.
To assess the effectiveness of our framework, we evaluate the social visual behaviors of the children in different therapy settings via (1) mutual gaze ratio generated by our framework; and (2) human-coded ratio (see Table~\ref{tab:t2}).
By mean comparison, the results show that the mutual gaze ratio was higher in the Play Therapy group ($M\!=\!0.118$, $SD\!=\!0.079$) compared to the Standard Therapy group ($M\!=\!0.101$, $SD\!=\!0.058$).
This positive trend for the Play Therapy group aligns well with the result on the human-coded ratio, which also shows a higher ratio for the Play Therapy group ($M\!=\!0.285$, $SD\!=\!0.171$) than the Standard Therapy group ($M\!=\!0.193$, $SD\!=\!0.162$).
However, the difference between the two group settings is not significant for mutual gaze ratio ($p\!=\!0.539$, $t_{(26)}\!=\!-0.622$, $ns$---$ns$ stands for statistically non-significant) or human-coded ratio ($p\!=\!0.182$, $t_{(26)}\!=\!-1.374$,$ns$) based on the independent $t$-tests.

\begin{table*}[t]
    \footnotesize
    \caption{Summary of mutual gaze ratio, duration, and human-coded ratio in different therapy groups and activities.}
    \label{tab:t2}
    \centering
    \resizebox{\textwidth}{!}{
    \begin{tabular}{llcccc}
    \toprule
    \multirow{2}{*}{Group} & \multirow{2}{*}{Activity} & \multirow{2}{*}[1ex]{\# of Observation} & \multirow{2}{*}[1ex]{Mutual Gaze Duration} & \multirow{2}{*}[1ex]{Mutual Gaze Ratio} & \multirow{2}{*}[1ex]{Human-coded Ratio}\\
    & & \multirow{2}{*}[1ex]{($N$)} & \multirow{2}{*}[1ex]{($M \pm SD$)} & \multirow{2}{*}[1ex]{($M \pm SD$)} & \multirow{2}{*}[1ex]{($M \pm SD$)}\\\midrule
     \multirow{3}{*}{Play Therapy} & Hello Song & $7$ & $53.40 \pm 12.02$ & $0.166 \pm 0.095$ & $0.403 \pm 0.191$ \\
     & Music Making & $11$ & $57.44 \pm 10.44$ & $0.088 \pm 0.052$ & $0.210 \pm 0.109$ \\
     & Combined & $18$ & $55.82 \pm 10.87$ & $0.118 \pm 0.079$ & $0.285 \pm 0.171$ \\
     \midrule
    Standard Therapy & Reading & $10$ & $61.07 \pm 11.45$ & $0.101 \pm 0.058$ & $0.193 \pm 0.162$ \\\midrule 
    Total & & $28$ & $57.92 \pm 11.18$ & $0.112 \pm 0.072$ & $0.252 \pm 0.171$\\
    \bottomrule
    \end{tabular}}
    
        
        
    \begin{center}
        {\raggedright
        \textit{
        The mutual gaze ratio from the framework and the human-coded ratio from the human coder are both in the [0--1] range. \\        
        Mutual gaze duration is the mean value of the frame numbers when the child is maintaining a social gaze behavior over $25$ frames ($1$ second).}
        \par}
    \end{center}
\end{table*}

\begin{figure*}[t]
    \centering
     \centering
    \begin{subfigure}[t]{0.33\textwidth}
        \centering
        \includegraphics[width=0.98\textwidth]{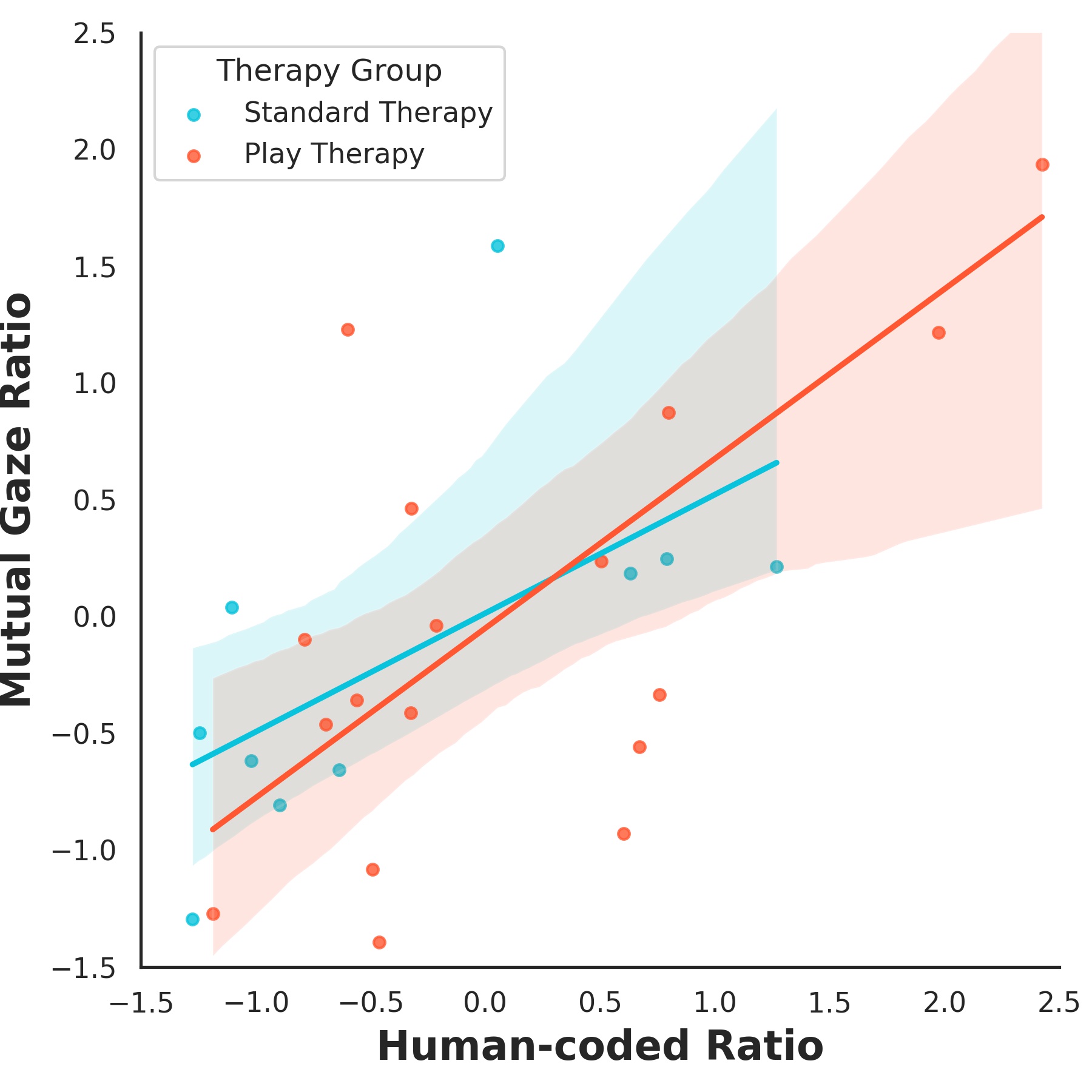}
        \caption{}
        \label{fig:scatter}
    \end{subfigure}
    \hfill
    \begin{subfigure}[t]{0.66\textwidth}
        \centering
        \includegraphics[width=0.98\textwidth]{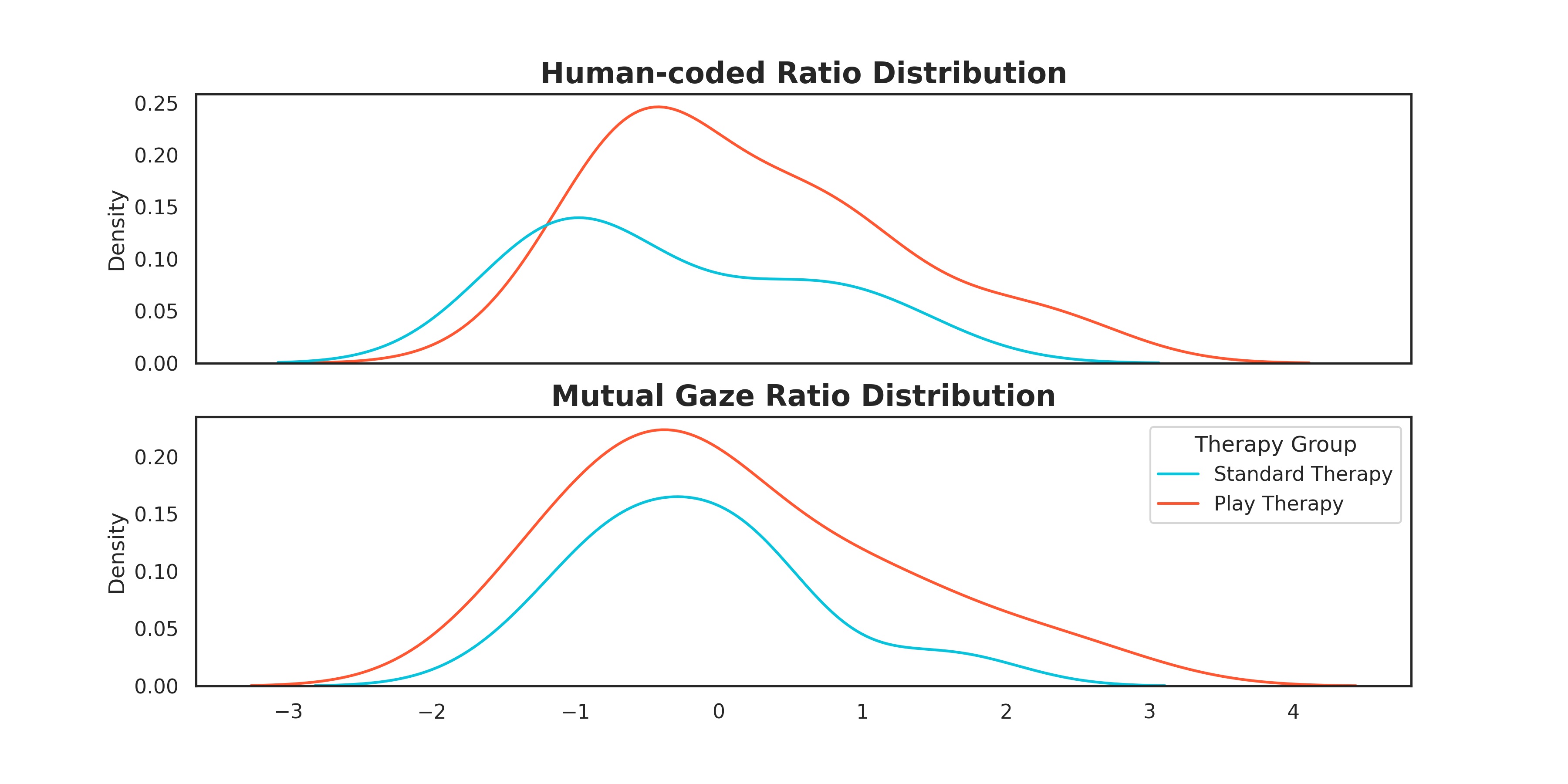}
        \caption{}
        \label{fig:distribution}
    \end{subfigure}
    \caption{(a) The scatter plot of mutual gaze ratio and human-coded ratio from $\mathbf{28}$ observations in different therapy group settings. The regression line and $\mathbf{95\%}$ confidence interval (shaped area) for each group are also included. (b) The distributions of mutual gaze ratio and human-coded ratio in different therapy group settings using kernel density estimation. The distributions of the two ratios are very similar for both therapy group settings. Both ratios are standardized.}
    \label{fig:correlation}
\end{figure*}

We further investigated the association between the mutual gaze ratio and human-coded ratio measures by calculating the Pearson correlation coefficient~\cite{benesty2009pearson}. 
The result indicated a strong positive linear relationship between the mutual gaze ratio and the human-coded ratio with statistical significance ($F_{(1,26)}\!=\!18.33$, $r_{p}\!=\!0.643$, $p\!<\!0.0005$, $\mathrm{RMSE}\!=\!0.056$).
The linear relationship and distributions of the standardized mutual gaze ratio and human-coded ratio are shown in Figure~\ref{fig:correlation}. 
As the similarity of these two distributions highlights, we can suggest that the mutual gaze ratio is a reliable indicator of social visual behavior. 
Thus, it shows the effectiveness of our mutual gaze detection framework as an automated assessment tool for evaluating and representing the social behavior of autism.

\subsection{Social Visual Behavior Analytics for Children with Autism}

We analyzed the social visual behavior of the children across therapy settings (Play Therapy and Standard Therapy), within-group activities (''Music Making'' activity and ''Hello Song'' Activity in the Play Therapy group), and therapy sessions (early and late sessions) using the measures described in Section \ref{Sec:Measures}.

\paragraph{\textbf{Social Visual Behavior in Different Therapy Settings}}

\begin{figure*}[t]
    \centering
    \includegraphics[width=0.8\textwidth]{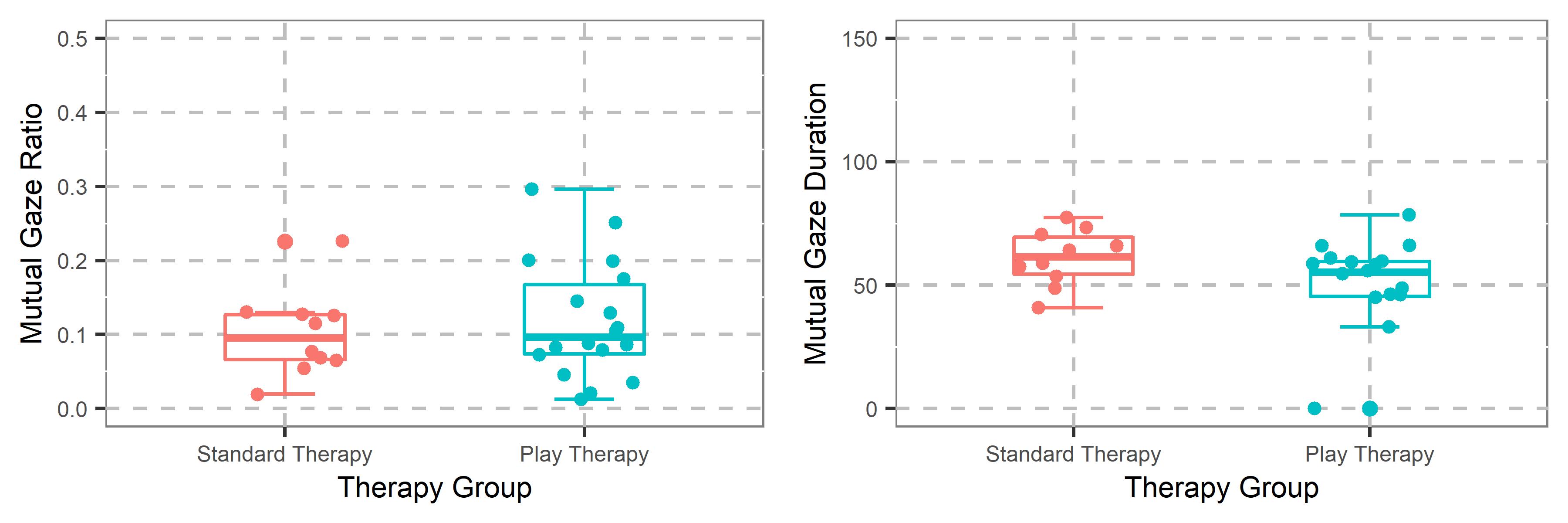}
    \caption{The box plots of mutual gaze ratio and mutual gaze duration in the Play Therapy group and the Standard Therapy group. No significant between-group difference in the mutual gaze ratio or duration. The mutual gaze ratio is in [$\mathbf{0}$--$\mathbf{1}$] range.}
    \label{fig:scatter1}
\end{figure*}

Independent $t$-tests were used to compare the mutual gaze performance (mutual gaze ratio and mutual gaze duration) of children between different therapy groups.
Although the children in the Play Therapy group got a higher mutual gaze ratio ($M\!=\!0.118$, $SD\!=\!0.079$) than the Standard Therapy group ($M\!=\!0.193$, $SD\!=\!0.162$), there were no significant between-group difference on the mutual gaze ratio ($p\!=\!0.539$, $t_{(26)}\!=\!-0.622$, $ns$)
, and neither on the mutual gaze duration 
($p\!=\!0.259$, $t_{(23)}\!=\!1.157$, $ns$).
Figure~\ref{fig:scatter1} shows the box plots of mutual gaze ratio and mutual gaze duration in different group settings.

\paragraph{\textbf{Social Visual Behavior During Different Activities in the Play Therapy Group}}

\begin{figure*}[t]
    \centering
    \includegraphics[width=0.8\textwidth]{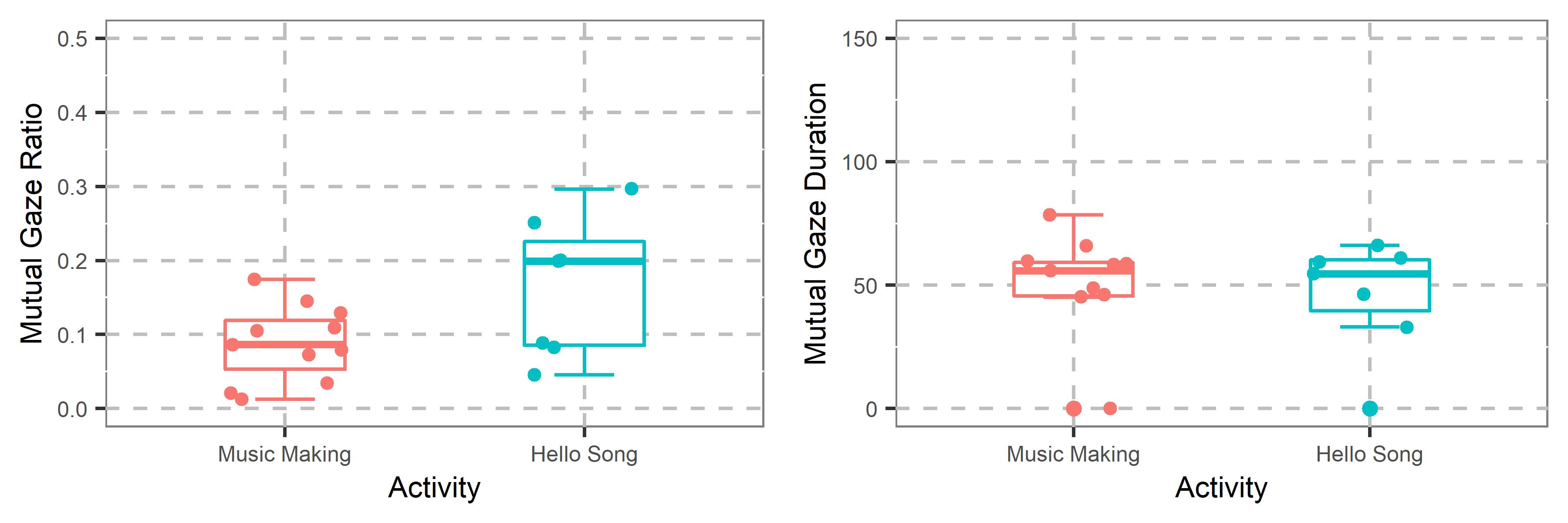}
    \caption{The box plots of mutual gaze ratio and mutual gaze duration in the ''Hello Song'' and the ''Music Making'' activities within the Play Therapy group. The mutual gaze ratio in the ''Hello Song'' activity is significantly higher than the ratio in the ''Music Making'' activity. No significant within-group difference in mutual gaze duration. Mutual gaze ratio is in [$\mathbf{0}$--$\mathbf{1}$] range.}
    \label{fig:scatter2}
\end{figure*}

To investigate the effects of within-group activities, independent $t$-tests were used to compare the children's mutual gaze performance between the ''Music Making'' activity and the ''Hello Song'' activity in the Play Therapy group.
There was a statistically significant within-group difference in the mutual gaze ratio ($p\!<\!0.05$, $t_{(16)}\!=\!-2.28$, $Hedge's\:g\!=\!-1.05$)
between the ''Hello Song'' activity ($M\!=\!0.166$, $SD\!=\!0.095$) and the ''Music Making'' activity ($M\!=\!0.088$, $SD\!=\!0.052$) with a large effect size (see Figure~\ref{fig:scatter2}).
On the contrary, no significant within-group difference was found on the mutual gaze duration ($p\!=\!0.501$, $t_{(13)}\!=\!0.693$, $ns$)
between the ''Hello Song'' activity ($M\!=\!53.40$, $SD\!=\!12.02$) and the ''Music Making'' activity ($M\!=\!57.44$, $SD\!=\!10.44$).
The results showed that different types of activities provided varieties of within-group effects on the frequency of the children's social gaze behaviors. 
Children performed gaze interaction with the trainers more frequently in the ''Hello Song'' activity than in the ''Music Making'' activity.

\paragraph{\textbf{Social Visual Behavior Improvement Across the Therapy Sessions}}

\begin{figure*}[t]
    \centering
    \includegraphics[width=0.6\textwidth]{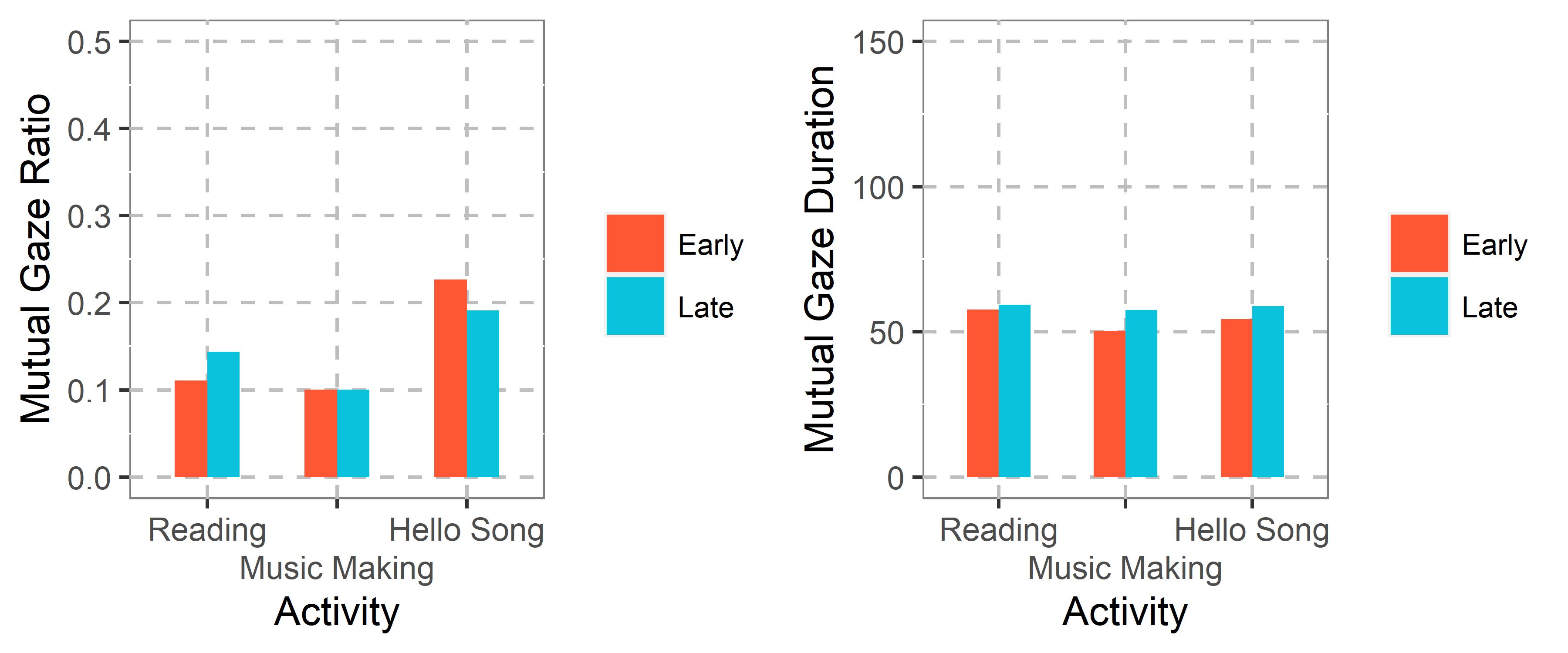}
    \caption{The bar chart of mutual gaze ratio and mutual gaze duration (frames) in different activities across early and late sessions. No significant social gaze improvement was found. Mutual gaze ratio is in [$\mathbf{0}$--$\mathbf{1}$] range.}
    \label{fig:barchart}
\end{figure*}

Descriptive and inferential statistics results about mutual gaze measures in different therapy groups and activities across therapy sessions were reported in Table~\ref{tab:t3} and Figure~\ref{fig:barchart}.
Independent $t$-tests were used to compare the mutual gaze performance (mutual gaze ratio and mutual gaze duration) of children across the early and late sessions (session $1$ and $16$).
Results showed that neither mutual gaze ratio ($p\!=\!0.928$, $t_{(54)}\!=\!-0.090$,$ns$)
nor social visual duration ($p\!=\!0.523$, $t_{(37)}\!=\!-0.645$,$ns$)
had significant improvement after the therapy training process.
However, compared to the mutual gaze performance in the early session, $1$ observation in the Hello Song activity, $4$ observations in the Music Making activity, and $5$ observations in the Reading activity had higher mutual gaze ratios and longer duration in the late session.

\begin{table*}[t]
    \footnotesize
    \caption{Summary of mutual gaze ratio and duration in different therapy groups and activities across therapy sessions.}
    \label{tab:t3}
    \centering
    \begin{tabular}{llcccc}
    \toprule
    & & \multicolumn{2}{c}{Play Therapy} & Standard Therapy& \multirow{2}{*}{Total}\\
    & & Music Making & Hello Song & Reading & \\\midrule
    \multirow{2}{*}{\# of Observation ($N$)} & \multirow{2}{*}[1ex]{Early Session} & $11$ & $7$ & $10$ & $28$\\
     & \multirow{2}{*}[1ex]{Late Session} & $11$ & $7$ & $10$ & $28$ \\\midrule
     \multirow{2}{*}{Mutual Gaze Ratio ($M \pm SD$)} & \multirow{2}{*}[1ex]{Early Session} & $0.100 \pm 0.102$ & $0.227 \pm 0.130$ & $0.111 \pm 0.134$ & $0.136 \pm 0.128$\\
     & \multirow{2}{*}[1ex]{Late Session} & $0.100 \pm 0.137$ & $0.191 \pm 0.129$ & $0.144 \pm 0.098$ & $0.139 \pm 0.123$ \\\midrule
     \multirow{2}{*}{Mutual Gaze Duration ($M \pm SD$)} & \multirow{2}{*}[1ex]{Early Session} & $50.26 \pm 19.31$ & $54.32 \pm 11.69$ & $57.74 \pm 26.86$ & $54.61 \pm 22.04$ \\
     & \multirow{2}{*}[1ex]{Late Session} & $57.46 \pm 13.28$ & $58.89 \pm 29.02$ & $59.28 \pm 12.64$ & $58.62 \pm 16.15$ \\
    \bottomrule
    \end{tabular}
    \begin{center}
        {
        \centering
        \textit{Early and late sessions represent the therapy training session $1$ and $16$.
        The mutual gaze ratio is in the [0--1] range.
        \\
        Mutual gaze duration is the mean value of the frame numbers when the child is maintaining a social gaze behavior over $1$ second ($25$ frames). 
        \\
        Neither mutual gaze ratio nor duration had significant improvement after the therapy training process.}
        \par}
    \end{center}
\end{table*}

\subsection{ Social Visual Behavior Score Prediction Using Machine Learning Models}

We evaluated the performance of the social visual behavior score (manually scored by human experts) prediction based on our mutual gaze-related measures (mutual gaze ratio and mutual gaze duration) and the participant profile information.
We particularly employed two profile data: level of functioning and ADOS-2 social affect score, which are directly related to the children's cognitive ability and diagnosis of ASD (see Table~\ref{tab:t1} for the details).
To investigate the benefit of our mutual gaze ratio measures as critical features to predict the social visual behavior score, we conducted an ablation experiment by comparing the prediction performance between two settings: (1) prediction based on our model with the mutual gaze ratio and duration from the framework together with the level of functioning and ADOS-2 social affect score, and (2) prediction only based on those two profile scores. We use the following five regression models: random forest (RF~\cite{breiman2001random}), support vector regression (SVR~\cite{drucker1996support}), Lasso regression (Lasso~\cite{tibshirani1996regression}), gradient boosting trees regression (GBT~\cite{schapire2003boosting}), and multi-layer perceptron regression (MLP~\cite{hornik1989multilayer}). 
Mutual gaze ratio, mutual gaze duration, level of functioning, and ADOS-2 social affect score are normalized.
To mitigate the training issues due to the small sample size, we used the Bootstrap approach to quantitatively control and check the stability of the results by random sampling with replacement. 

We report the $R^{2}$, root mean squared error ($\mathrm{RMSE}$), and mean absolute error ($\mathrm{MAE}$) for each predictive model in each setting (see Table~\ref{tab:t4}).
According to the results, predictions with mutual gaze-related measures yield a lower loss than the predictions only based on participant profile data in all models.
The random forest regression model with mutual gaze-related measures achieved the best prediction performance ($\mathrm{MAE}\!=\!0.348, \mathrm{RMSE}\!=\!0.447, R^{2}\!=\!0.620$).

\begin{table}[t]
    \footnotesize
    \caption{Summary of social visual behavior score prediction model performance in the ablation experiment.}
    \label{tab:t4}
    \centering
    \begin{tabular}{lcccccc}
    \toprule
     & \multicolumn{3}{c}{Prediction Setting 1} & \multicolumn{3}{c}{Prediction Setting 2}\\
    Model & $\mathrm{MAE}$ & $\mathrm{RMSE}$ & $R^{2}$ & $\mathrm{MAE}$ & $\mathrm{RMSE}$ & $R^{2}$ \\\midrule
    \textbf{RF} & \textbf{0.348} & \textbf{0.447} & \textbf{0.620} & $0.386$ & $0.464$ & $0.578$\\
    Lasso & $0.395$ & $0.486$ & $0.558$ & $0.408$ & $0.505$ & $0.494$\\
    SVM & $0.428$ & $0.513$ & $0.493$ & $0.449$ & $0.530$ & $0.481$\\
    GBT & $0.455$ & $0.566$ & $0.373$ & $0.482$ & $0.596$ & $0.369$\\
    MLP & $0.538$ & $0.623$ & $0.241$ & $0.595$ & $0.671$ & $0.159$\\
    \bottomrule
    \end{tabular}
    \begin{center}
        {\raggedright
        \textit{
        Prediction Setting 1: prediction based on the mutual gaze ratio and duration, together with the level of functioning and ADOS-2 social affect score.\\        
        Prediction Setting 2: prediction only based on the two profile scores, including the level of functioning and ADOS-2 social affect score. \\    
        Mutual gaze ratio, mutual gaze duration, level of functioning, and ADOS-2 social affect score are normalized. \\   
        The prediction model with setting 1 which includes gaze information has better performance versus the prediction model in setting 2 without gaze information (lower values are better for $\mathrm{MAE}$ and $\mathrm{RMSE}$, and higher value is better for $R^{2}$).}
        \par}
    \end{center}
\end{table}

\section{Discussion}
\label{Sec:Discussion}

Here we proposed a social visual behavior analytical approach for children with autism using mutual gaze detection.
We detected the mutual gaze behaviors by adopting a state-of-the-art gaze detection method~\cite{9311655} and generated mutual gaze-related measures, mutual gaze ratio, and mutual gaze duration, based on the detection outcomes.
We compared the outcomes of the framework with a manually-coded ground truth measure of social gaze (human-coded ratio) to evaluate the effectiveness of our framework and analyzed the children's social visual behavior using mutual gaze-related measures with an ASD dataset collected from the autism therapy intervention.

As reported in the previous section, our results show that our framework successfully reports a mutual gaze ratio, which is comparable, even replaceable to the human-coded ratio.
For different therapy settings, descriptive comparison results of our method match the results of human-coded measures, which were annotated by subject matter experts.
We also found a strong positive linear correlation between the mutual gaze ratio and the human-coded ratio. 
As opposed to traditional hand-annotating approaches, our method is able to extract mutual gaze features effectively and efficiently throughout the entire training process with a single standard camera.

During the social visual behavior analysis, group setting differences, activity-related within-group effects, and intervention-related changes in social attention were examined.
The results show that there is no significant difference in children's social attention frequency or duration in different group settings.
However, in the same group, different activities have significant effects on the frequency of children's social gaze behaviors.
Children in the Play Therapy group performed mutual gaze behavior more frequently during the ''Hello Song'' activity than the ''Music Making'' activity.
This may be caused by the fact that children with autism focus more on objects used in training, including instruments such as drums and xylophones, rather than humans~\cite{thorup2017gaze}.
Given the easy access to objects, children may have engaged in a greater visual fixation on objects.
Moreover, the practice of complex drumming and xylophone patterns required sustained monitoring of the instruments.

These findings are consistent with the results of the ''Reading'' activity, in which children with autism also focus on the books when they are reading.
Thus, as many researchers report, object-based activities may reduce children's social visual behaviors during the autism training process~\cite{scassellati2007social,lian2022influence,vacas2021visual}.
After comparing the mutual gaze ratio and duration between early and late sessions, no significant improvement in the social visual performance of children with autism was found after the therapy sessions. 
This is consistent with the findings of Srinivasan \emph{et al.}~\cite{srinivasan2016effects} that assessed social gaze for the entire session for early and late sessions based on data coded by human experts. They too reported activity-related differences with great socially-directed gaze during musical play therapy sessions. 
As acknowledged in the previous studies, it is very challenging to change levels of sustained social attention in children through short-term behavioral interventions~\cite{hassan2021serious,leung2021effectiveness}.

By examining the prediction models with and without mutual gaze-related measures, we found that the models with mutual gaze ratio and duration have better prediction performance than the models only based on participant profile data. 
In this case, our mutual gaze-related measures derived from the mutual gaze detection framework can be an important feature to predict the social visual behavior score and be used for social visual behavior evaluation. 
Among five different regression models, the random forest regression model with our mutual gaze-related measures can provide the best prediction performance with the lowest loss.
Since the social gaze behaviors are scored based on various social gaze patterns, to reduce the loss, more social gaze patterns need to be collected, measured, and used for predictions.

\paragraph{Limitations and future work}

The work reported here has certain limitations that we would like to emphasize and expand on in the future.
The framework~\cite{9311655} we adopted in this paper detected the mutual gaze features based on head tracks, which may not be able to capture the correct gaze information for specific situations, \emph{e.g.}, when the eye gaze is not matched with the head direction,\emph{e.g.}, children with ASD could be using peripheral vision and non-foveal vision for tracking individuals; when the head is occluded; when the video record has a low resolution or zoom in/out capturing.
To overcome this issue, we plan to investigate and develop more accurate and robust gaze detection methods.
We also consider reducing the bias in measurement, annotation, and sampling~\cite{tay2022conceptual}.
This work is still a preliminary study with a limited number of samples.
Like most autism studies, this sample was biased toward males given the greater ASD prevalence in males.
Given the small sample size of this study, we acknowledge that the effect sizes calculated in this study can be imprecise with large confidence intervals.
Despite the absence of social visual behavior annotations, there is still a large subset of collected data that needs to be considered in our future work.
Further evaluation and discussion with clinical experts would be invaluable in developing more effective gaze detection, and social behavior evaluation approaches.
In the future, we aim to improve our framework via semi-supervised and unsupervised machine learning methods to analyze those video recordings with minimum ground truth annotations~\cite{emerson2020early,tay2022conceptual}. 
Moreover, we plan to investigate multimodal, multi-gaze features, including gaze sharing and following~\cite{lian2018believe}, facial expressions~\cite{minaee2021deep,li2021two}, engagement \cite{zhu2020MultiRate} and body gestures~\cite{heredia2021automatic,li2022dyadic,li2022pose} for framework development and comprehensive interpretation of social visual behaviors.

\section{Conclusion}
\label{Sec:Conclusion}

Observing children's gaze interactions during therapeutic training sessions, particularly their mutual gazes toward trainers, can provide valuable social visual information; therefore, analyzing such behaviors is extremely useful for treating and evaluating children effectively.
In this paper, we analyzed the social visual behavior performance of the children with autism from the ASD dataset using advanced mutual gaze detection framework~\cite{9311655}.
The effectiveness of the framework was validated by comparing the outcome with the social visual behavior measure hand-coded by subject matter experts.
According to the social visual behavior analysis, we found that the social attention patterns of children may be affected by certain contexts (with or without learning tools) during short-term behavioral interventions.

The reported ablation experiment showed that mutual gaze measures could be a powerful feature for social visual behavior score prediction.
The creation and extraction of mutual gaze features can provide valuable information for more accurate analysis and prediction model development.
Our findings have implications for social interaction analysis in education technology, therapy evaluation, and intervention design, by offering an analytical approach with a novel assessment tool for social behaviors based on gaze information.
Beyond the autism therapy context, our method can be applied to other application scenarios that need reliable automatic social behavior analysis from videos involving human-human or human-robot interactions.

\bibliographystyle{ACM-Reference-Format}
\bibliography{main}

\end{document}